\documentclass[useAMS,letterpaper]{mn2e}
\usepackage{graphicx}
\usepackage{latexsym}
\usepackage{amsmath}
\usepackage{amssymb}
\usepackage{subfigure}
\usepackage{url}

\title[Stellar collisions in accreting protoclusters]{Stellar collisions in accreting protoclusters: a Monte Carlo dynamical study} 

\author[O. Davis, C.  Clarke \& M.  Freitag]{O. Davis $^{1,2}$\thanks{email: olaf.davis@astro.ox.ac.uk}, C. J. Clarke$^{1}$ \& M. Freitag $^{1,3}$\\
$^1$Institute of Astronomy, Madingley Road, Cambridge, CB3 0HA, UK\\
$^2$Astrophysics, Keble Road, Oxford, OX1 3RH, UK\\
$^3$Gymnase de Nyon, Route de Divonne 8, 1260 Nyon, Switzerland}

\begin{document}

\pagerange{\pageref{firstpage}--\pageref{lastpage}} \pubyear{2010}

\newcommand{\dd}{\textrm{d}}
\newcommand{\rin}{R_\textrm{\tiny{in}}}
\newcommand{\OO}{\mathcal{O}}
\maketitle

\label{firstpage}

\def\mnras{MNRAS}
\def\apj{ApJ}
\def\aap{A\&A}
\def\apjl{ApJL}
\def\apjs{ApJS}
\def\araa{ARA\&A}

\begin{abstract}
We explore the behaviour of accreting protoclusters with a Monte Carlo
dynamical code in order to evaluate the relative roles of accretion,
two body relaxation and stellar collisions in the cluster evolution.
We corroborate the suggestion of Clarke \& Bonnell that the number
of stellar collisions should scale as $N^{5/3} \dot M^{2/3}$ (independent
of other cluster parameters, where $N$ is the number of stars in the cluster
and $\dot M$ the rate of mass accretion) and thus strengthen the argument
that stellar collisions are more likely in populous (large N) clusters.
We however find that the estimates of Clarke \& Bonnell were pessimistic
in the sense that we find that more than $99 \%$ of the stellar collisions
occur within the post-adiabatic regime as the cluster evolves towards
core collapse, driven by a combination of accretion and two-body
relaxation. We discuss how the inclusion of binaries may reduce the
number of collisions through the reversal of core collapse but also note
that it opens up another collisional channel involving  
the merger of stars within 
hard binaries; 
future Nbody simulations are however required in order to explore this
issue.
\end{abstract}


 \section{Introduction}
 
  Bonnell, Bate \& Vine (1998) first pointed out  that star clusters
should shrink, in response to adiabatic accretion of gas onto their
stars, and conducted Nbody experiments to illustrate this process.
They also hypothesised that, in the process, cluster cores should
be driven to such high densities that the stars there could 
physically collide, and 
proposed this  as a formation mechanism for massive stars.
An attractive feature of this scenario is that it avoids a well
known problem with conventional (accretion) models for 
star formation which predict that  the effect of radiation pressure  
on dust in the accretion flow can be problematical once stars exceed around
$10 M_\odot$. It is likely, however, that the original
arguments against the formation of massive stars by accretion 
are most severe
in spherical geometry (Wolfire \& Casinelli 1987) and pilot simulations
demonstrate that high stellar masses  ($> 10 M_\odot$)
are attainable through accretion 
in more realistic (disc) geometry (Yorke \& Sonnhalter 2002). The
recent identification of flattened structures around high mass protostars
(Cesaroni et al 2006) lends further credence to the idea that accretion is a viable
formation mechanism for many massive stars. The question that remains
unanswered  - and which this paper attempts to explore - is  whether 
there are ever astrophysical environments in which one
would  expect the collisional formation
of massive stars to be important? 

 Following the original suggestion of Bonnell et al 1998,  
Bonnell \& Bate (2002) pursued hydrodynamical simulations of 
the shrinkage of cluster cores due
to mass loading 
(in this case modeling 
accretion, from gas initially within the cluster, onto a dynamically
cold population
of cluster stars). They found however, that they were unable to pursue
the calculation into the regime of physical collisions unless the
nominal radius of each star was increased by an order of magnitude.
More recently, Clarke \& Bonnell (2008)  presented physical arguments
as to what limits the maximum density attainable in clusters subject to
adiabatic accretion and used these arguments to try and define the
parameters of accreting clusters in which stellar collisions might be expected. 

  In this paper we conduct a further study of  accretion onto
stellar clusters, using a Monte Carlo dynamical code.  A Monte Carlo
code is well suited to this problem since it follows the secular evolution
of clusters that are subject to processes occurring over many dynamical times,
rather than following the  trajectories of individual stars in their
orbits. In the present case, the two (secular) processes that are modeled
are a) the addition of gas onto the stars and b) the diffusion of stellar
orbits as a result of two-body relaxation. The latter process is modeled
by random selection and perturbation of neighbouring particles as described
in Freitag \& Benz (2001,2002) and in Section 2 below. The mean (smooth) potential of the cluster is
however spherically symmetric so that gravitational forces are easily
computed from the enclosed mass. This means that the code avoids the
(potentially $\propto N^2$) expense of calculating pairwise forces over
N particles and hence can readily be used for integrating systems of much
higher N than is possible via  direct Nbody integration. This is particularly
useful in the present problem where our initial estimates (Clarke \& Bonnell
2008) suggested that stellar collisions are only likely to occur in
relatively high N systems.

It should however be stressed at the outset 
that this study is highly idealised in a number of respects: i) it
treats a spherical cluster of equal mass stars which all accrete zero
momentum gas at a constant rate,
ii) it omits binaries and three body dynamical interactions and iii) it
assumes the gaseous mass reservoir originates outside the cluster and that
$100 \%$ of the gas accreted ends up on the stars, rather than as
a distributed gaseous component within the cluster (see Bonnell et al 1998
for a discussion of the case where the infalling gas also forms a distributed
component). Each of the above simplifications can be relaxed in future
studies, building on previous research into non-accreting clusters;
the goal of the present work is to set out an analytic framework
for understanding this simplest case (verified through numerical simulations)
and to use this to further explore how the incidence of stellar 
collisions depends on cluster parameters. We note that since the simulations
described here can be uniquely specified by two dimensionless numbers
(the initial ratios of the two-body relaxation and collision time-scales to
the mass accretion time-scale),
it is a simple matter to apply the results to other physical situations
(e.g. a cluster of black holes intercepting a spherical accretion flow).

  The structure of the paper is as follows. In Section 2 we describe
the evolution of accreting proto-clusters, clarifying the expected
properties in the adiabatic regime and  explaining the circumstances under
which clusters embark on adiabatic shrinkage and how this is modified
by two-body relaxation. In Section 3 we briefly describe the Monte Carlo
code and the simulations undertaken. Section 4 discusses the results
and Section 5 contains some estimates of how these results would be
likely to be modified if the dynamical role of binary heating had been
included. Section 6 presents conclusions and indicates further necessary
research directions. 

\section{ Outline of the problem} 
\label{sec-outline}

The rate of collisions per unit volume in a stellar system is given by

\begin{equation}
\label{eq-cdot}
\dot{c} = 8\sqrt{\pi}G\frac{n^2M_\star R_\star}{v}\left(1+\frac{2v^2R_\star}{GM_\star}\right),
\end{equation}

\noindent{}where $M_\star$ and $R_\star$ are the stellar mass and radius,
$v$ is the velocity dispersion and $n$ the number density (Binney \& Tremaine 1988).

  Collisional formation of massive stars during the pre-main sequence stage implies
a value for $\dot{c}$ of $\gg1$ per Myr. For physically realistic $M_\star$, $R_\star$
(i.e. on the order of solar values)
and a velocity dispersion on the order of 10km/s, this 
 requires that clusters attain (albeit briefly) extremely high densities
of around $10^8$ stars pc$^{-3}$. This value exceeds those encountered in
even the  densest regions of observed Galactic clusters (such as the
Arches Cluster: see Table 4 in Stolte et al 2006) by more than two orders
of magnitude and argues that such dense conditions, if ever encountered,
must represent a short-lived evolutionary stage. Other effects that
would render such dense stellar concentrations hard to observe include
heavy extinction (if the cluster were  still deeply embedded in its natal
gas) or the fact that an ultra-dense core may be too small to resolve
in distant clusters (Trager et al 1995). 

A mechanism for producing 
such ultra-dense conditions  (Bonnell et al 1998) 
involves the shrinkage of a cluster in response to the accretion
of material onto its constituent stars. 
If one can neglect evolution due to two-body dynamical effects (see below)
then the only agent of evolutionary change is the cluster's response to mass
gain. In the limit that the time-scale
for mass doubling (henceforth $t_{\dot M} = M/\dot M$) is long
compared with the cluster dynamical time ($t_{dyn}$) this mass gain
is {\it adiabatic} and in this case adiabatic invariance requires
that the cluster radius shrinks according to $R \propto M^{-3}$ 
in response to mass gain. 
{\footnote{ This result can be readily derived
in several ways: either by considering the (conserved) angular momentum
of individual particle orbits or else by considering the re-virialisation
of a cluster whose kinetic energy is reduced, and magnitude of potential
energy increased, through the  incremental addition of zero momentum
gas. Note
in the latter derivation  that mass  must be added
incrementally (i.e. a small mass gain per dynamical time) in order to
attain the strong shrinkage predicted - in the limit that mass is
added instantaneously, by contrast, revirialisation only results
in shrinkage by up to a factor two, however much mass is added.}}  

 Adiabatic shrinkage under mass accretion thus results in a dramatic
increase in cluster density:   the number density scales as
$M^9$ while the mass density as $M^{10}$. It thus needs to be borne
in mind that  extreme conditions can result from
relatively small (order unity) changes in the cluster mass. The over-all
time-scale for shrinkage (and possible production of a core where
collisions are important) is just $\sim t_{\dot M}$ as defined above.
It will also be useful to note that the cluster velocity dispersion
rises more mildly with mass (i.e. as $M^2$) and we therefore find that
stellar encounters generally remain in the gravitationally focused regime
even when the cluster is dense enough for collisions to occur. 

  The adiabatic requirement ($t_{dyn} < t_{\dot M}$) can be recast as
an upper limit on the rate of accretion onto a cluster, i.e.
$\dot M < v^3/G$ where $v$ is the cluster velocity dispersion. Although
we use the term `cluster' to describe our stellar system throughout this
paper, we  note that the  shrinking adiabatic system
can in fact be the core of a larger cluster,  in which  gas accretes
radially on to the core from a reservoir in the outer cluster.
In the case that free falling  gas dominates the mass of the  outer cluster,
the 
adiabatic condition requires that the velocity
dispersion in the core be larger than the free-fall velocity of the
inflowing gas.   Once this (not implausible) condition is attained, the
core can decouple as an adiabatically shrinking sub-system. 

  In Section 4.2 and below  we discuss the range of initial cluster parameters and accretion
rates that fall in the adiabatic regime. The steep rise in density
with added mass, and the consequent reduction in the dynamical time-scale,
means that any cluster which satisfies the adiabatic condition
at the onset of shrinkage will satisfy this condition ever more amply
as the collapse proceeds.   However, the reduction of the dynamical
time is associated with a proportional reduction in the two-body
relaxation time-scale (which, to within a logarithmic factor, scales
as $N$ crossing times) and so,  at some point in the collapse,  two-body
relaxation is likely to occur on a time-scale that is comparable with
the shrinkage time-scale $t_{\dot M}$. From this point on it is not
possible to regard the cluster evolution as being collisionless;
Clarke \& Bonnell made the (pessimistic) assumption that this evolutionary
point marks the end of the era of opportunity for stellar collisions,
since they argued that thereafter the core would be inflated by two-body
scattering. The viability of collisions in this case depends on the 
number of collisions that are possible before this point; evidently this
favours high N clusters since the relative smoothness of the potential
implies that such clusters can collapse by a large factor before two-body
effects become important. 

 In what follows, we explore this issue with a Monte Carlo dynamical
code with the aim a) of refining our initial estimates of how
far the cluster can collapse in the adiabatic regime and b)
pursuing the evolution into the subsequent regime that is driven
by two-body relaxation.  

\section{The simulations}

 \subsection{The Monte Carlo dynamical code} 
 
  All simulations described here are performed using ME(SSY)**2,
a Monte Carlo dynamical code that is fully described in Freitag
\& Benz (2001, 2002) and Freitag (2008).
{\footnote{The code is available online at http://www.ast.cam.ac. uk/research/repository/freitag/MODEST\_MonteCarlo/MESSY\_ Download.html.}}
In this code the cluster is approximated as a 
set of spherical shells of zero thickness (henceforth termed `particles') 
which represent a layer of one or more stars that share the same stellar
properties (mass) , orbital parameters (kinetic energy and angular momentum)
 and phase. The phase and radial position of this particle in its
orbit is not continuously evolved, since the particle does not represent
a particular set of stars but instead the set of stars that at any time
have this particular set of orbital properties. The cluster is set up
(and always remains) in a state of dynamical equilibrium so that
in the absence of the perturbations described below, the particles
conserve their individual angular momenta and energies. The code only
follows the evolution that results from changes on  time-scales that are  
much longer than a typical stellar orbital period.

 One of these processes is diffusive relaxation whereby the orbital
properties represented by Monte Carlo particles are slowly perturbed
in a way that mimics the cumulative effect of small angle deflections
from other stars. This is implemented by randomly selecting 
pairs of nearby particles and effecting a two body encounter between
them that changes their orbital properties ( kinetic
energy and angular momentum).  
Each is then given a random phase and placed in a corresponding position on
its orbit. The justification for this is that  
several dynamical times will have elapsed before two-body interactions 
have any effect on the overall cluster, so that the particle  has time to
`forget' its previous phase.

  The other process included in the code is the accretion of gas; in
these simplest calculations, all particles share the same mass
at a given time and gain mass at a uniform rate. Since the
accreted gas is assumed to have zero momentum, the kinetic energy of
each particle is reduced appropriately by gas accretion. As particles
change their masses, kinetic energies and angular momenta as
a result of accretion and diffusive relaxation it is necessary to
re-assign their radii within the cluster and hence to take account
of secular changes in the smooth component of the cluster potential.
This latter is easily computed given the assumption of spherical
symmetry. Monte Carlo codes thus generically avoid the problem
encountered by direct Nbody integration codes which require the 
computation of the potential and force resulting from $N^2$
pair-wise forces. Consequently Monte Carlo codes represent
a highly computationally efficient method for following the 
evolution of large N systems over a large number of dynamical times.

 \subsection{ Simulation properties} 

    We model a cluster as a spherical system of equal mass point masses 
distributed according to a Plummer law, i.e. with density profile
$\rho \propto (1+r^2/a^2)^{-5/2}$ (see e.g. Binney \& Tremaine 1988). 
The initial isotropic velocity dispersion is scaled
so as to ensure a state of virial equilibrium; the Monte Carlo code
enforces a state of global virial equilibrium throughout, as is to
be expected in the case of adiabatic accretion (i.e. accretion on
a time-scale much longer than the dynamical time-scale). The only two
evolutionary processes followed by the code are those of
mass accretion and two body relaxation. Thus, a given simulation can be
scaled so as to describe an entire family of clusters and accretion rates,
provided  that they have the  same initial ratio of the mass accretion time-scale
to the two body relaxation time-scale. In addition, however, we trace the
incidence of stellar collisions, through post-processing of the densities
and velocity dispersions yielded by the Monte Carlo code. 
It should be emphasised,
however, that the collisions are {\it not} implemented in the code, but merely
recorded. The incidence of collisions breaks the degeneracy between models
with the same initial ratio of two body relaxation time-scale to mass accretion
time-scale, and introduces a second dimensionless number, the ratio of the
initial collision timescale to the mass accretion time-scale.

\section {Results} 

\begin{figure}
\centering
\resizebox{\columnwidth}{!}{\includegraphics*{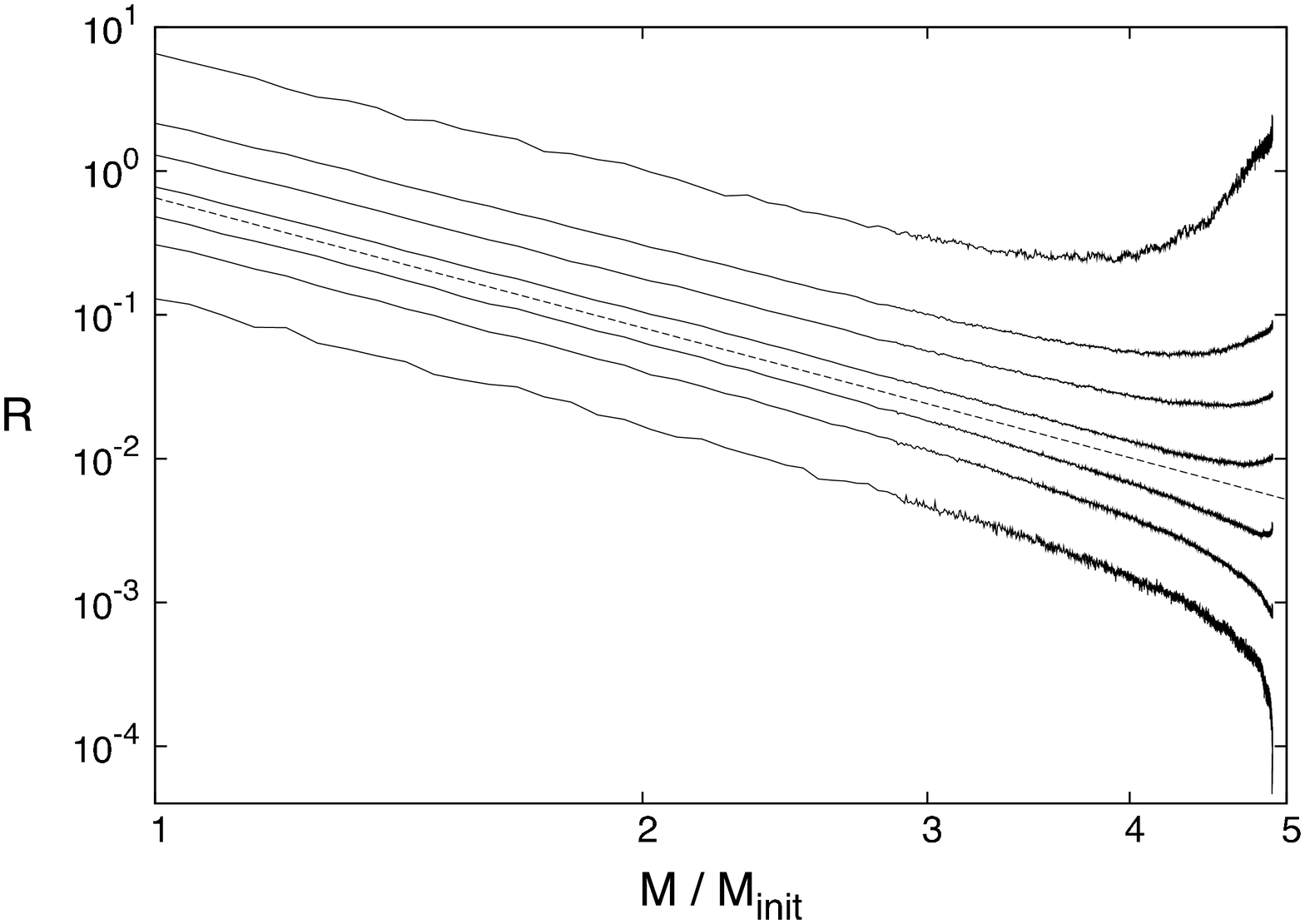}}
\caption{Evolution of various Lagrangian radii (containing $0.99,0.90, 0.75,0.5,0.25,0.1$ and $0.01$ of the cluster mass from top to bottom) as a function
of the total cluster mass for a simulation with $\gamma_{init}=140$ (see equation (\ref{eq-gamma})). The dots are $R\propto M^{-3}$.}§
\end{figure}

\subsection{ Cluster density evolution} 

  Figure 1   depicts the evolution of various Lagrange radii (each individually
tracing the location of a point containing a fixed {\it fraction} of the
cluster mass) as a function of the total cluster mass. This plot illustrates
the two key phases of cluster evolution. At early stages of the collapse,
the Lagrange radii track the expected result for adiabatic accretion,
i.e. $R \propto M^{-3}$: the cluster evolution is self-similar, so that
the initial Plummer sphere is mapped onto successively more compact Plummer
distributions as the mass increases. However at later times (when the
mass has increased by a factor $\sim 4$ for this particular simulation)
deviations from self-similarity become apparent: the inner Lagrange radii
start to decline more rapidly, while the decline of larger Lagrange radii
starts to flatten out and then increase again. This latter behaviour may be
readily understood as being a consequence of two-body relaxation. The 
diffusion terms in the Monte Carlo code have the effect of transferring
energy outwards through the cluster, and correspond, in a real cluster,
to the integrated effects of two body encounters. (This inexorable outward
transfer of energy is a consequence of the negative heat capacity of
self-gravitating systems such that energy loss from the core, and
associated shrinkage and revirialisation, results in an {\it increase}
in the core's kinetic energy (`temperature'). This   then drives a further
departure from thermal equilibrium; this process is sometimes termed
the `gravothermal catastrophe'.) The ultimate consequence
of such transfer, in a cluster consisting of single stars, is `core collapse'
whereby the central regions attain nominally infinite densities (see
discussion in Section 5 of the role of
three-body effects in averting this outcome). The
time-scales for the latter stages of core collapse are short and thus
negligible mass is accreted during this stage (hence the near vertical
evolution in Figure 1 at late times). To first order, then, the evolution
can be divided into an adiabatic regime (where two body effects are negligible)
followed by a core collapse regime, wherein accretion is negligible.
Needless to say,  this is an over-simplification and the Monte
Carlo code  is needed to follow the transition between these regimes
correctly. Nevertheless, we will use such a simple conceptual framework in
order to find analytical scalings for collision frequencies, for comparison
with our numerical data. Finally, we note that the over-all evolutionary
time-scale is set by the time-scale for mass addition, $t_{\dot M}$; since the
density is such a steep function of total mass in the adiabatic regime
($\rho \propto M^{10}$), the end of the adiabatic regime/core collapse
is achieved during the time that the cluster mass increases only by
a factor of order unity. The mass doubling time-scale, $t_{\dot M}$  thus
sets the over-all evolutionary clock.

  All the clusters modeled have a qualitatively similar evolution to that
shown in Figure 1. What however does depend on initial conditions is the
collapse factor  in the adiabatic regime (see Figure 2). 
One can  derive this dependence 
on cluster parameters by exploiting the   self-similar nature of the collapse
in the adiabatic regime. We expect two-body relaxation to cause 
significant deviations from adiabatic collapse at the mass scale  ($M_a$)
at which
the two-body relaxation time-scale, $t_{2r}$, becomes comparable with 
the mass doubling time-scale and have verified numerically
that this is indeed the case. The former is given by:

\begin{equation}
t_{2r} = {\frac{N}{8\rm{ln} \Lambda}} \sqrt{{\frac{R^3}{GM}}} 
\end{equation}

\noindent{}where $R$ is the cluster radius and  ln$\Lambda$ is the Coulomb
logarithm, where $\Lambda = \gamma_c N$ and $\gamma_c \sim 0.1$ for
a system of single masses (Freitag et al 2006). The mass doubling time-scale 
is 
given by

\begin{equation}
\label{eq-tmdot}
t_{\dot M} = {\frac{M}{\dot M}}
\end{equation}

Thus, since $R$ shrinks as  $M^{-3}$ during the evolution
and we are considering the cases where both
$\dot M$ and $N$ are constant, we obtain

\begin{equation}
t_{\dot M} \propto M
\end{equation}

\noindent{}and 

\begin{equation}
t_{2r} \propto M^{-5}   
\end{equation}

  Thus if we define the initial ratio of these time-scales as $\gamma_{init}$: 

\begin{equation}
\label{eq-gamma}
\gamma_{init} =   {{t_{2r}}\over{t_{\dot M}}}|_{init}
\end{equation}

\noindent{}then we expect the ratio, $\gamma$ to evolve as:

\begin{equation}
\gamma = \gamma_{init} \bigl({{M}\over{M_{init}}}\bigr)^{-6}
\end{equation}

  Consequently we expect the mass scale where $\gamma = 1$ (i.e.
mass shrinkage and two body relaxation occur on the same time-scale) to
satisfy:

\begin{equation}
\label{eq-gamma1/6}
{{M_{\gamma=1}}\over{M_{init}}} \sim \gamma_{init}^{1/6}
\end{equation}

\begin{figure}
\centering
\resizebox{\columnwidth}{!}{\includegraphics*[angle=270]{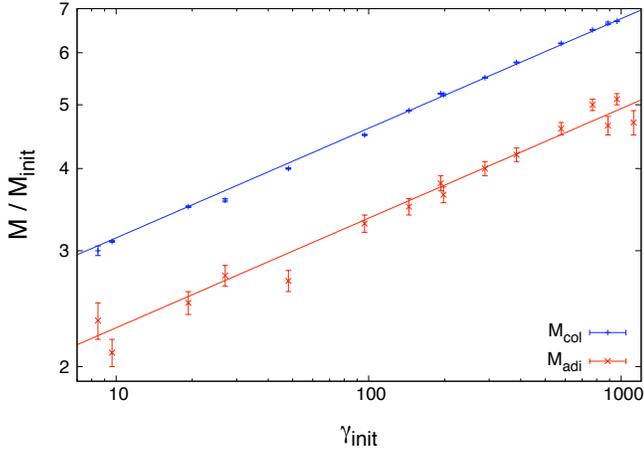}}
\caption{The ratio of cluster mass to initial cluster mass corresponding
to a) the end of adiabatic collapse (lower points)  as estimated from
the evolution of the Lagrangian radii and b) the point
of core collapse (upper points), both  as a function of the initial value of
$\gamma$ (equation (\ref{eq-gamma})), the ratio of the two-body relaxation time-scale to
the mass doubling time-scale. The errorbars represent our uncertainty
in deriving these quantities visually from the simulation output; the relatively 
low scatter suggests that this method of identifying the end of adiabaticity is 
sufficient for present purposes.}
\end{figure}

  In Figure 2 we plot two mass scales as a function of $\gamma_{init}$:
 the lower points denote
estimates of the mass scale at which adiabatic collapse ends (based on visual
estimates of where the Lagrangian radii start to deviate significantly
from the $R \propto M^{-3}$ scaling) while
the upper points instead represent the mass gained at the point 
of core collapse.
 Both quantities are parallel to the line with slope
$1/6$ and therefore correspond to lines of constant $\gamma$ (equation (\ref{eq-gamma1/6})).
Note that the lower points correspond to  $\gamma \sim 0.1$:
this is consistent with a picture in which two body effects first
become significant at $\gamma \sim 1$ but where, in common with purely
stellar dynamical calculations (e.g. Freitag \& Benz 2001), 
this does not start to manifestly
affect the evolution of the Lagrangian radii until several relaxation
time-scales have elapsed.    

 We again note the rather modest
mass gains that are required even to reach the point of core collapse
and that the majority  (around $2/3$) of the
mass is added to the cluster in the adiabatic
regime. Nevertheless  the fact that the remaining 
$1/3$ of the  mass is added 
after the end of the adiabatic regime
emphasises that the
latter phases of the cluster evolution are driven by a mixture of
accretion and two body effects and that it is only in the final stages
of core collapse that it becomes an essentially stellar dynamical
problem. 

\subsection{ Collisions in the adiabatic regime} 

 It is straightforward to show, given the scalings of velocity and
density with mass given above, together with an expression for
the collision rate in the gravitationally focused regime (i.e. equation (\ref{eq-cdot}) when dominated by the first term in brackets), that the
number of collisions per mass doubling time-scale scales as $M^{10}$.
{\footnote{Strictly the scaling is as $(M^{10} - M_\text{init}^{10})$, 
but the former term quickly comes to dominate with increasing $M$, as shown by Figure~8.}}
Since we have shown above that the mass added in the
adiabatic regime scales as $\gamma_{init}^{1/6}$ it then follows
that the total number of collisions in the adiabatic regime is
$\sim C_0 \gamma_{init}^{5/3}$, where $C_0$ is the number of
collisions expected in the first mass doubling time, i.e. the
product of the initial collision rate ($\dot C_0$) and
$t_{\dot M}$. In the gravitationally focused regime the collision cross
section 
scales as the product of the gravitational radius ($Gm_*/v^2$ where
$m_*$ is the stellar mass) and  the stellar radius ($r_*$), so that the
collision rate per star is the product of this cross section with 
the number density and velocity dispersion;  the total collision
rate is $N$ times this value. Putting all this together we find
$\dot C_0 \propto N^2 m_* r_*/R^3v$ and $C_0 \sim \dot C_0 t_{\dot M}
\propto N^2 M  m_* r_*/(R^3v \dot M)$ 
 We also see that
$\gamma_{init}$ scales as $N \dot M/v^3$ and thus that the
total number of collisions in the adiabatic regime ($C_a$) scales
as $N^{11/3}  M \dot M^{2/3}  m_* r_*/R^3 v^6$. Applying the
virial condition ($v^2 \propto M/R$ and $M = N m_*$) we then find
that we are left with the predicted scaling 

\begin{equation}
\label{eq-cadi}
C_a \propto
N^{5/3} \dot M^{2/3} r_*/m_*.
\end{equation} 

Although $m_*$ and $r_*$ both
change during the simulation, we assume that their ratio is
invariant (although in higher mass stars, the stellar radius is
a rather weaker function of stellar mass).
Thus we are left with the result that $C_a$ depends on only two
parameters ($N$ and $\dot M$): notably neither  the total cluster mass
nor radius appear in this expression. The lower set of points in
Figure 3 illustrate that this predicted scaling with $\dot M$ and
$N$ is indeed reproduced by the simulations. 

\begin{figure}
\centering
\resizebox{\columnwidth}{!}{\includegraphics*[angle=-90]{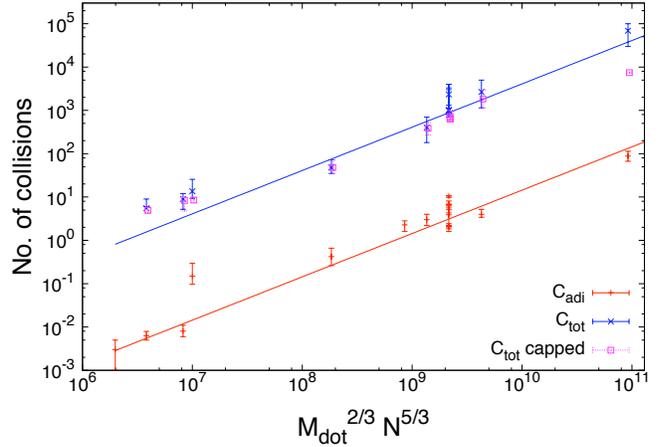}}
\caption{The total number of collisions as a function of the quantity
$\dot M^{2/3} N^{5/3}$, in $(M_\odot/$Myr)$^{2/3}$. Lower points: collisions in the
adiabatic regime. Upper points: total number of collisions up to
core collapse. The magenta and blue points correspond to capped and uncapped
values respectively (see text)}
\end{figure}

  Having established this scaling, it is instructive now to plot the
total number of collisions in the adiabatic regime in the plane of
accretion rate versus $N$, since (for fixed $m_*/r_*$) these
are the only cluster parameters that affect $C_a$. Such a plot is shown
in Figure 4. The crosses represent
simulations annotated by the total number of collisions in the adiabatic
regime as obtained by post-processing (note that some points here correspond to
more than one point on Figure 2, since fixing $\dot{M}$ and $N$ does not fix
$\gamma$; in these cases the label shows average collision number). The lines represent contours of
constant predicted $C_a$ with numerical values fit to the results
of Figure 3. Figure 4 shows that in general these fits are quite
good predictors of the total number of collisions in the adiabatic
regime. Figure 4 also indicates a couple of restrictions on input
parameters. We have already noted (Section \ref{sec-outline}) that the adiabatic condition
implies that the initial cluster velocity dispersion $v$ is such
that $v^3 > G \dot M$. The minimum value of $v$ as a function
of $\dot M$ is shown on the top horizontal axis. These minimum
velocity dispersions are rather low and rise only very mildly with
$\dot M$; thus we see that clusters with realistic velocity dispersions
($>$ a few km/s) will undergo adiabatic shrinkage even in the
case that $\dot M$ is as high as $10^5 M_\odot/$Myr.

\begin{figure}
\centering
\resizebox{\columnwidth}{!}{\includegraphics*[angle=-90]{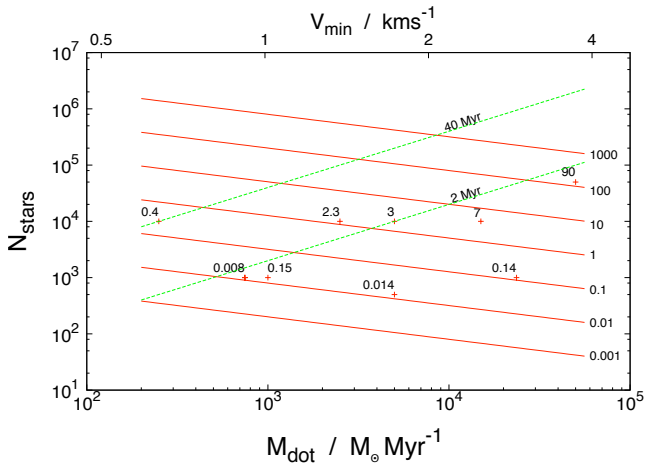}}
\caption{Contours of equal numbers of collisions in the adiabatic regime
as a function of $\dot M$ and $N$ obtained by fitting the lower points
in Figure 3. The datapoints show the individual values of $N$, $\dot{M}$
for the simulations which were performed; attached numbers show the
mean number of collisions for simulations at that point.
The rising dashed lines are contours of constant
mass doubling time-scale ($t_{\dot M}$). The upper horizontal axis indicates
the minimum $v$ required to satisfy $v^3 > G \dot M$ for given $\dot{M}$ 
(and hence to ensure that the mass doubling
timescale exceeds the dynamical timescale: see Section 2).}
\end{figure}

  Figure 4 also shows two lines such that the over-all system evolution
time-scale ($t_{\dot M}$) is $2$ and $40$ Myr, on the assumption that the cluster
initially consists of stars of mass $1 M_\odot$. These two time-scales bracket
the main sequence lifetimes of stars that are supernova progenitors and 
producers of powerful winds. If star formation in the cluster (in the
absence of collisions) leads to stars in this mass range then at some
point in the range $2-40$ Myr we expect supernova/stellar wind feedback to
halt accretion and thus (unless the cluster is already well on the
way to core collapse) further cluster shrinkage would halt at that
point. The positions of these lines therefore imply that rapid cluster
shrinkage (occurring on less than the time-scale for feedback to be effective)
is associated with the lower right hand part of the diagram.
 We therefore see that significant numbers of collisions in the
adiabatic regime are only produced in clusters that are
vigorously accreting at the uncomfortably high rate of $> 10^{4} M_\odot$ Myr$^{-1}$ (which corresponds to 
of order a per cent of the entire galactic star
formation rate). If such conditions were ever realised,  the optimal
cluster scale for producing a significant number of collisions in a short
time would be  $N \sim 10^4$.

 \subsection{Collisions  in the post-adiabatic regime: the path to  core collapse} 

 \begin{figure}
\centering
\resizebox{\columnwidth}{!}{\includegraphics*[angle=-90]{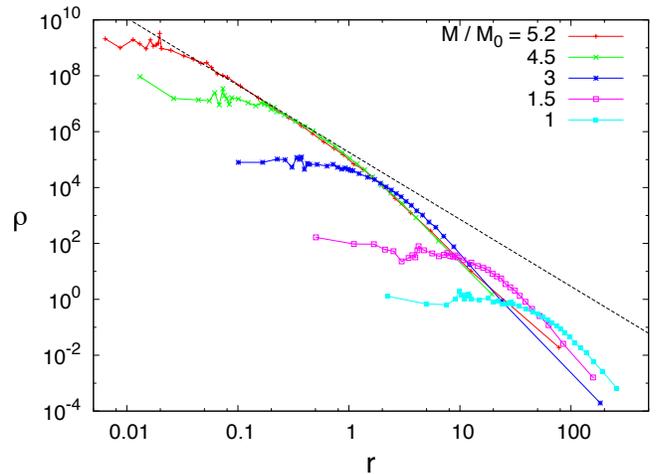}}
\caption{Evolution of the cluster density profile during collapse. 
The dashed line depicts the relationship $\rho \propto r^{-2.4}$ as
is expected to develop outside the core on the way to core collapse. 
Units are arbitrary.}
\end{figure}

  Figure 5 depicts snapshots of the cluster density profile: the 
right hand profile  represents the initial Plummer sphere and
the next two profiles are essentially scaled copies of this profile,
illustrating that, as expected, the cluster evolves homologously in the
adiabatic regime. The left hand two profiles however illustrate the
development of a power law profile at intermediate radii with a slope
of $\rho \propto r^{-2.4}$. This structure is characteristic of
the density profile expected during core collapse (see Lynden-Bell
\& Eggleton
1980 for analytical arguments in favour of this
power law profile outside the cluster core). Figure 6 is
a schematic depiction of the evolution of the profile during core collapse:
the point separating the `core' {\footnote {Note that the `core' that
develops during core collapse should not be confused with the `core'
of the initial Plummer profile.}} and `halo' (regimes A and B)
evolves along the line shown: the core density rises during the
approach to core collapse ($\rho_c \propto r_c^{-2.4}$) but 
the fraction of mass contained in the
core ($\propto r_c^{0.6}$) falls during this period.

\begin{figure}
\centering
\resizebox{\columnwidth}{!}{\includegraphics*[angle=-90]{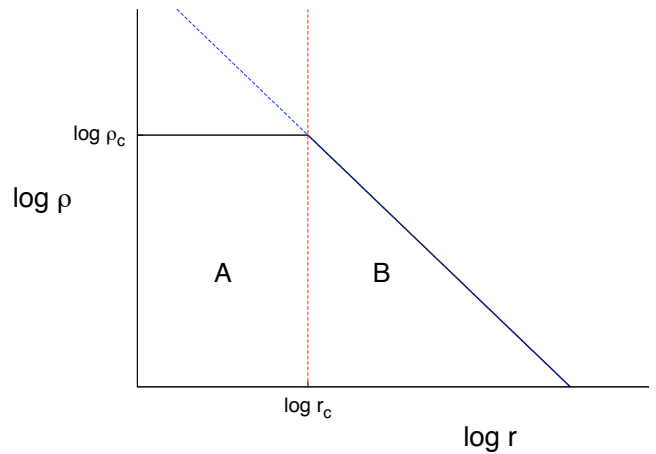}}
\caption{Schematic depiction of the cluster density profile en route to
core collapse. The core radius and density ($r_c$ and $\rho_c$) evolve along 
the diagonal line with $\rho \propto r^{-2.4}$.}
\end{figure}

  This behaviour  turns out to  be an important factor
in limiting the total number of collisions during core collapse
to finite values.  We can estimate whether we expect this quantity to be
convergent by noting that the remaining time to core collapse
($\tau$) can be expressed as a multiple $\eta^{-1}$ of the instantaneous
two body relaxation time-scale as core collapse is approached. Given the
scaling of two body relaxation time-scale with the core mass and radius ($r_c$)
and given the evolution of the core density along the locus $\rho_c
\propto r_c^{-2.4}$ (Figure 6) we can show that $r_c \propto (\tau \eta) ^{5/9}$. We can also show that the total number of collisions during core collapse
is dominated by those occurring in the vicinity of $r_c$ and that the
total collision rate scales as $\dot C \propto r_c^{-1.6}$. This
then implies that $\dot C$ scales as $(\tau \eta)^{-8/9}$. Integrating
this with respect to $\tau$ one sees that the predicted total number
of collisions that are yet to occur in the remaining time ($\tau$)
to core collapse is a (weak) positive power of $\tau$ - i.e. it is not
expected to diverge as core collapse approaches (provided that $\eta$
does not fall indefinitely as core collapse is approached). This is consistent with
our finding that the total number of collisions recorded is not arbitrarily
sensitive to the fineness of the time sampling in the approach to
core collapse. Nevertheless it  is worth noting that the collision rate
is strongly peaked towards the final phases of core collapse. 

\begin{figure}
\centering
\resizebox{\columnwidth}{!}{\includegraphics*[angle=270]{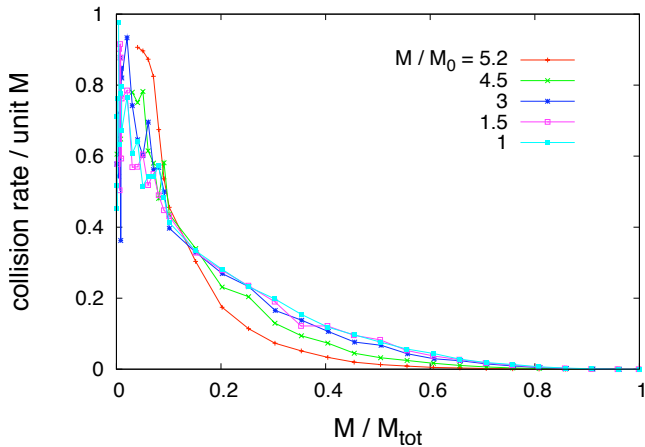}}
\caption{Collision rates per unit mass as a function
of the fractional mass enclosed, for the same five times shown in Figure 5.
The y-axis is in arbitrary units and is normalised such that the fraction of collisions occurring within
a given Lagrangian mass interval at a particular time is simply proportional
to the area under the graph. The radial profile of collisions is centrally concentrated at all times, 
becoming increasingly so during the final stages of core collapse  (more than
$80 \%$ of the collisions occurring in the innermost $20 \%$ of the mass).}

\end{figure}

  Figure 7 depicts the spatial location of where collisions
are expected to occur, with the area under the curve depicting the total
number of collisions occurring in any given mass range. Evidently,
most of the collisions occur in the inner $10-20 \%$ of the mass, which
is consistent with the fact that it is only the Lagrangian radii enclosing
less than $\sim 10-20 \%$ of the mass that are involved in core collapse
(see Figure 1). This concentration of collisions within a small mass
fraction of the cluster has prompted us to examine whether the total
number of collisions recorded are independent collision events or
whether they represent multiple collisions of a given set of stars.
In what follows we sometimes refer to `capped' collision rates which have
been obtained by limiting the total number of collisions in a Lagrangian
shell to the number of stars in that shell. 
Simulations for which the uncapped rate is
significantly larger than the capped rate are those where we expect
a collisional runaway to be occurring in the core. The actual numbers
become particularly unreliable once one enters a regime where stars
are typically undergoing multiple collisions given 
that we at no
stage feed back into the calculation the effects of any collisions that
are expected to occur.  Nevertheless the existence of simulations where
the capped and uncapped values are quite different is a useful indicator
of the regions of parameter space where we expect 
multiple collisions to be occurring.   

  The upper points in Figure 3 represent the (capped and uncapped) numbers
of collisions at the point of core collapse for the simulations for
which the lower points denote the number of collisions during the adiabatic
regime. We see that the total number of collisions follows exactly the
same ($\propto \dot M^{2/3} N^{5/3}$) scaling as the number of collisions
in the adiabatic regime,  but offset by a factor of about $300$. We have
already noted that,  in the adiabatic regime, the number of
collisions should scale as $M^{10}$ and that the mass increases by
a further $\sim 1.5$ during post-adiabatic evolution. Thus our
finding that the total number of collisions in the post-adiabatic
regime (en route to core collapse) is around $300$ times higher
than in the adiabatic regime is {\it roughly} consistent with an
extrapolation of the number of adiabatic collisions to take account
of the extra mass accreted. The reason for this can be discerned from
Figure 1, which is a simulation in which the outer Lagrangian
radii start to deviate from self-similar (adiabatic) shrinkage at a mass 
gain of factor $\sim 3.5$ whereas
core collapse occurs at a mass gain factor of $\sim 5$. We see that 
for most of the intervening interval, the deviation of the
inner Lagrangian radii (where
most of the collisions occur) from the adiabatic tracks is rather mild
and 
only shows 
an abrupt downturn at the very end. The consequences of this evolution for the
collision history is depicted in Figure 8;  although the end of the
adiabatic regime (defined as where  the 
Lagrangian radii first start to deviate from adiabatic evolution)
occurs at a mass
gain factor of $\sim 3.5$, the total number of collisions continues to
rise as $\sim M^{10}$ until a mass gain factor of $>4$ (where $\gamma \sim 0.01$): 
at this point there is
a further rise in the number of collisions by about a factor three in
the last stages of core collapse.

  From an empirical point of view, therefore, we see that the  previous
assumption of Clarke \& Bonnell (2008) that collisions only
occur in the adiabatic regime was pessimistic; such collisions
only represent  $< 0.5\%$ of the total number expected in the
case that the cluster evolves all the way to core collapse. This is
shown in Figure 9, which is similar to Figure 4 but with an over-all
increase in the number of collisions by a factor $\sim 300$ (the figures
recorded in Figure 9 are uncapped; we see from Figure 3 that the
capped and uncapped numbers start to diverge  - indicating a regime
of multiple collisions locally - at the stage that the total number of
collisions is greater than a few hundred). This factor $300$ makes
a critical difference to our assessment of whether there are
environments in which we expect an astrophysically interesting number of
collisions. We see from Figure 9 that we now expect many tens of collisions
in clusters numbering more than a few thousand stars and with accretion 
rates exceeding $\sim 10^{3} M_\odot$/Myr. Both the threshold accretion rate
and the threshold $N$ values are an order of magnitude lower than in the case that one counts only
collisions in the adiabatic regime; {\footnote{As one would expect given the scaling of
the number of collisions with $\dot M$ and $N$ (equation (\ref{eq-cadi})):  we can use
equation (\ref{eq-tmdot}) (assuming that $M \propto N$) to show that the accretion rate
at which, say, there are $10$ collisions within $2$ Myr scales as
the $-3/7$ power of the boost factor ($300$ in this case). Thus the
required accretion rates and values of $N$ fall from $10^4 M_\odot$ Myr$^{-1}$
and $N=10^4$ to $10^3 M_\odot$ Myr$^{-1}$ and $N=1000$ respectively}} 
in particular, the required accretion rate is now more astronomically
plausible.

\begin{figure}
\centering
\resizebox{\columnwidth}{!}{\includegraphics*[angle=-90]{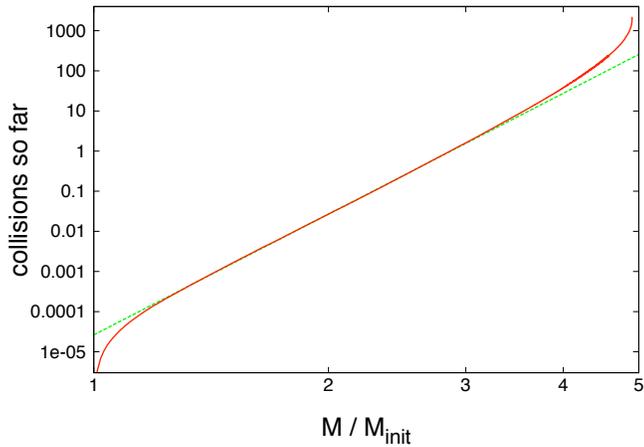}}
\caption{The total number of collisions as a function of cluster mass (normalised
to initial mass) for a simulation with $\dot M = 2000 M_\odot$ Myr$^{-1}, N=10^4$.
The number of collisions remains relatively close to the $M^{10}$ adiabatic scaling 
beyond the
point at which the Lagrangian radii start to deviate from homologous adiabatic
collapse (at $M/M_{init} \sim 3.5$) and then increases by a further factor $2-3$ in the final stages of core collapse.} 
\end{figure}

\begin{figure}
\centering
\resizebox{\columnwidth}{!}{\includegraphics*[angle=-90]{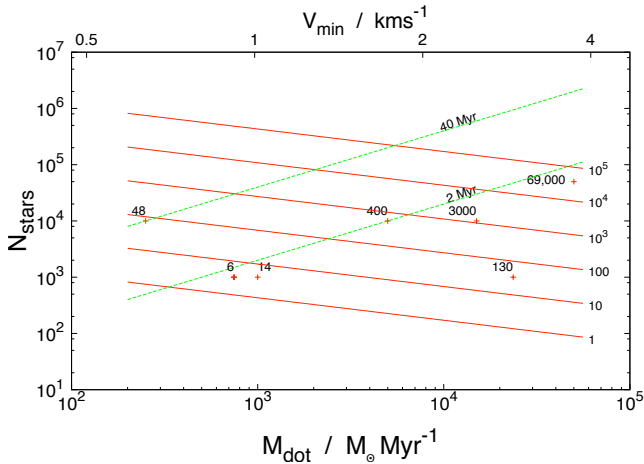}}
\caption{As Figure 4 but now for the total number of collisions up to the point
of core collapse. Note that these are uncapped collisions (see text) and not necessarily
physical in the larger cases. }
\end{figure}

 \section{  Averting core collapse and the role of binaries} 

  We have seen above that the generation of a large number of collisions
requires that the cluster shrinkage can proceed well into the post-adiabatic
regime. For the example shown in Figure 8, less than $1 \%$ of the total
collisions have occurred at the point that stellar dynamical relaxation effects
first become significant (i.e. $t_{2r} = t_{\dot M}$) and a further few $\%$ 
occur in a regime where the number of collisions continues to approximately follow the
same scaling ($\propto M^{10}$) as in the adiabatic regime. The rest of the
collisions occur in a regime where the  number of collisions then rises more
steeply as a function of added mass, with around a half to two thirds of the total occurring in the final moments of core collapse.
  
We however need to be cautious about necessarily concluding that a
real cluster will indeed follow this continued shrinkage of its inner
regions. Although the Monte Carlo simulations contain a correct
treatment of two-body relaxation (which is the driver of core collapse)
they omit the effect of binary stars in re-inflating the cluster core.
A well known effect in stellar dynamical simulations, which has come
to be known as Heggie's Law, is that the effect of interaction between hard
binaries (i.e. those with an orbital velocity exceeding the cluster
velocity dispersion) and other binaries and single stars is on average
to transfer energy from the binary to the stars' orbital motion
in the cluster potential. In a situation where stars in the
core are losing energy
by two-body scattering to dynamically colder stars at larger radius, their
consequent revirialisation in a more compact core configuration  (i.e
the process of core collapse) can be
offset by the re-injection of energy from binary interactions (see, for
example, the discussion in Chapter 9 of Heggie \& Hut 2003).  We therefore
need to ask whether, given a realistic binary population, core collapse
could be averted  - or, more precisely (given that we don't require core
collapse to go right to completion in order to produce a significant
number of stellar collisions) we need to ask at what stage in the collapse
will the core be re-inflated by the effect of binaries.

 In order to assess this issue, we will first consider the possible role
of three-body binaries (i.e. those that are dynamically created in the
cluster as a result of two stars becoming bound through
transference of energy to a third body). This process requires that
the three stars interact within an interaction volume of scale $Gm_*/v^2$
and is therefore favoured in small N systems where high stellar densities
do not also imply large velocity dispersions ($v$). Heggie \& Hut (2003)
estimate that core heating by three-body binaries is only important
in cores containing less than $\sim 50-80$ stars. Our cores contain $5-10 \%$
of the total number of stars in the cluster and so we only expect three-body
binaries to be important in rather low $N$ systems (i.e. those totalling
less than a few thousand stars). Since this is also the regime in which
we do not expect a significant number of collisions (Figure 9)  we
conclude that three-body binaries have little bearing on the incidence
of collisions in accreting clusters.

  We now turn to {\it primordial} binaries, i.e. those binaries that
are the product of the initial star formation process rather than
being a consequence of later dynamical interactions in the cluster.
We emphasise that we can only invoke the effect of {\it hard} binaries
for the re-inflation of the cluster core, since soft binaries instead
extract energy from the relative motion of stellar systems and are
an energy sink rather than energy source. Given the parameters of
the clusters  we are considering and the expected rise in the core velocity
dispersion en route to core collapse, the relevant hard-soft borderline for
binaries totalling a few solar masses is of order an A.U.. Therefore
it is only  binaries of sub- A.U. separation in the post-adiabatic
regime that can potentially avert core collapse. (Note, however, that
since the binaries themselves are subject to adiabatic shrinkage, these
hard binaries could evolve from binaries with initial separations of
up to $\sim 100$ A.U.).

There are several conditions that need to be met in order for primordial
binaries to be significant in offsetting core collapse but in the
present case the critical one is that  
they must be able to deliver energy to the core at the {\it rate} required
to counteract core collapse. The hard binary fraction ($f$) required to achieve
this is given (Heggie \&  Hut 2003) by:

\begin{equation}
f^2 = \eta {\rm log}\Lambda  
\end{equation}

\noindent{}where ${\rm log} \Lambda$ is the Coulomb logarithm and  $\eta$ is  the 
ratio of the two-body relaxation time-scale in the core to the
time remaining to core collapse ($\tau$). We have estimated $\eta $ 
from  the simulations and find that it is $\sim 0.003$ during the evolutionary
phase when most of the collisions occur, in agreement with simple analytic
estimates (Cohn 1980).
This implies that the hard binary fraction that is required in order to offset core collapse is of the order of
$15 \%$. This is within the range of realistic values for primordial fraction believed 
to apply in real clusters, suggesting that binaries may indeed play a significant role in some cases.

 We do however note that if such a population of hard binaries did exist,
{\footnote{Note that since we require $15 \%$ of all our stars
to be in hard binaries, the total energy contained in
hard binaries is of order the binding energy of the core  even if all
the hard binaries are only modestly (less than an order of magnitude)
harder than the hard-soft borderline. We would then satisfy another
condition for binaries to re-inflate the core (see Chapter 9 of  Heggie \& Hut 2003)}}
then one should also take into
account the {\it enhancement} in collisions due to dynamical hardening
of close pairs. Note that interactions that lead to mergers within
binary pairs require the third star to approach within about a binary
separation of the pair, as opposed to direct collisions which require
stars to approach to within a stellar diameter. The (gravitationally
focused) cross section for such interactions is thus much larger than for
direct collisions and it is thus unsurprising, in situations where there
are many hard binaries, that this should represent an important
collision channel (Bonnell \& Bate 2002, Gaburov et al 2008). Binary-binary interactions
are also an important source of collisions in binary-rich clusters, with
collisions due to such encounters dominating over those
involving single stars at a binary fraction ($\gtrsim 20\%$) fairly close to our
required value of $15\%$ (Bacon et al 1996).

 \section{Conclusions} 

 Our highly idealised exploration of the interplay between two-body
relaxation, adiabatic accretion and collisions in young star clusters
yields results that are qualitatively in line with the conclusions
of Clarke \& Bonnell (2008). The total number of stellar collisions
scales as $N^{5/3} \dot M^{2/3}$ (independent of cluster mass, radius
or velocity dispersion) and  is thus strongly favoured in
populous clusters.  We see from Figure 9  that stellar collisions are likely to be important
in populous clusters ($N \sim 10^4$ or more) provided that they are
subject to accretion on a reasonable~{\footnote{Here `reasonable' 
implies a mass doubling time-scale of $< 10^6$ years if cluster
shrinkage is required during the pre-main sequence stage; on the
other hand a mass doubling time-scale of $< 10^3$ years
would imply a super-Eddington
luminosity for the accretion of dusty gas onto the constituent stars
(Wolfire \& Cassinelli 1987). The
upper scale in Figure 9 indicates the minimum initial cluster velocity
dispersion that is required for adiabatic accretion at the rate given
by the corresponding figure on the lower scale. }} time-scale. 

  In contrast to the speculations of Clarke \& Bonnell, however, we find
that most ($\sim 99.7 \%$) of the collisions are expected to occur {\it
after} the point at which the two-body relaxation time-scale
becomes comparable with the mass doubling time. We find that the evolution
of Lagrangian radii starts to deviate from adiabatic collapse when
the ratio ($\gamma$) of two-body relaxation time-scale to mass doubling
time-scale is about $0.1$ but that the number of collisions continues
to grow as the $10$th power of the added mass (as in the initial
adiabatic phase) until $\gamma$ drops to $\sim 0.01$; the total number
of collisions rises by about an order of magnitude during this phase
when $\gamma$ is in the range $0.1$ to $0.01$. Thereafter, the cluster
undergoes the final stage of core collapse, boosting the total number
of collisions by a further factor $2-3$. 
We estimate that the shrinkage of the cluster core in our simulations
would
only be reversed 
if the cluster core contained
a large fraction ($\sim 15 \%$) of 
hard (sub A.U.) binaries and that in this case we might even find that
the collision frequency was enhanced by mergers of stars within dynamically
hardened binaries.
This is an issue that needs to  be
explored by Nbody models of accreting clusters which include a
significant hard binary population. 

Another significant simplification in this work was its adoption of
a single mass for all stars in the cluster, in contrast to realistic cluster
initial mass functions. The presence of a primordial IMF leads to mass
segregation as massive stars sink to the core through dynamical friction.
For realistic IMFs these massive stars form a dynamically decoupled population
at the cluster centre (Spitzer 1969).
This has the effect of shortening the relaxation time in the core
(e.g. G\"urkan et al 2004) and also increasing the collisional cross-section,
both of which can be expected to increase the collision rate; however,
a full understanding of how this will affect our results, particularly when
binaries are also involved, will require further investigation.

We have evidently arrived at the point where further progress requires
the relaxation of the large number of restrictive assumptions listed in
Section 1. This study has clarified the evolution of the idealised model
and demonstrated that future work needs to focus on the role of binaries
and mass segregation in the post-adiabatic regime.    

\section*{Acknowledgements}
The authors wish to thank H. Belkus for helpful feedback on the manuscript.
OD is grateful for a grant provided by the Institute of Astronomy during part of this work.

\label{lastpage}

\end{document}